# Cell Trapping Utilizing Insulator-based Dielectrophoresis in The Open-Top Microchannels

Chun-Ping Jen*, Yao-Hung Huang and Teng-Wen Chen
Department of Mechanical Engineering, National Chung Cheng University,
No. 168, University Rd., Min-Hsiung, Chia Yi, 62102, Taiwan, R.O.C.

*Abstract-* The ability to manipulate or separate a biological small particle, such as a living cell and embryo, is fundamental needed to many biological and medical applications. The insulator-based dielectrophoresis (iDEP) trapping is composed of conductless tetragon structures in micro-chip. In this study, a lower conductive material of photoresist was adopted as a structure in open-top microchannel instead of a metallic wire to squeeze the electric field in a conducting solution, therefore, creating a high field gradient with a local maximum. The microchip with the open-top microchannels was designed and fabricated herein. The insulator-based DEP trapping microchip with the open-top microchannels was designed and fabricated in this work. The cells trapped by DEP force could be further treated or cultured in the open-top microchannel; however, those trapped in the microchip with enclosed microchannels could not be proceeded easily.

## I. INTRODUCTION

The ability to manipulate particles, especially living cells, is essential to many biological and medical applications, including isolation and detection of rare cancer cells, concentration of cells from dilute suspensions, separation of cells according to specific properties, and trapping and positioning of individual cells for characterization. Several kinds of physical forces can be employed for particle manipulation. Those are optical [1], mechanical [2], magnetic [3,4], electrical [5,6], and other manipulations [7,8].

Dielectrophoresis (DEP) has been widely used for the manipulation, separation and characterization of cells, DNA, virus, and colloid particles (in both micro- and nano-scales) in various microfluidic platforms. DEP force is generated by electrode structures of appropriately designed such as the pin-plate [9] isomotive [10], polynomial [11], or castellated interdigitated forms [12]. Early studies of DEP used electrode structures made from thin metal wires, needles, or plates, while modern DEP employs microfabrication technology to produce microelectrode arrays capable of producing sufficiently large DEP forces to induce particle motion with small applied voltages. DEP on micro-fabricated electrodes has been proved especially suitable for its relative ease of micro-scale generation and structuring of an electric field on microchips. Moreover, integrated DEP biochips provide the advantages of speed, flexibility, controllability and ease of application to automation because DEP traps consist of scalable electrode arrays be designed to pattern thousands of cells on a single glass slide. DEP have been using as cell trapping [13], levitation [14], separation [15], and sorting [16] on an electrode array by varying electrode shape and arrangement. DEP on micro-fabricated electrodes has been proved especially suitable for its relative ease of micro-scale generation and structuring of an electric field on microchips.

DEP force field could also be generated by using dielectric constrictions or insulating obstacles, therefore, these methods further extend the scope of DEP applications. These approaches have been termed electrodeless dielectrophoresis (EDEP) [17] or insulator-based dielectrophoresis (iDEP) [18], respectively, in distinguishing from the method conventionally employed by metal electrodes-based DEP for force generation.

In this study, a lower conductive material of photoresist (JSR, THB 151N) was adopted as a structure in open-top microchannel instead of a metallic wire to squeeze the electric field in a conducting solution, therefore, creating a high field gradient with a local maximum. The microchip with the open-top microchannels was designed and fabricated herein. The schematic illustration of the trapping device is shown in Fig 1. This trapping microchip is suitable for trapping cells or biological samples and easily proceeding further treatment for cells, such as culturing or contact detection. Biological sample and buffer loading before analysis and cleaning after analysis in the proposed microchip with the open-top microchannels could become easier and faster, hence, it is beneficial in high-speed sequential multiple measurements. Furthermore, the Joule heat generated by applying high voltage in the open-top microchannel could dissipate more effectively than that in the enclosed microchannel, because that it is open to the air. When the fluorescent detection is used, the intensity of the emitted light will not be interfered by the cover, which is absent in the microchip with open-top microchannels.

## II. DIELECTROPHORETIC (DEP) FORCE THEORY

The phenomenon of DEP is first defined by Pohl (1978) as the motion of neutral particles caused by dielectric polarization effects in non-uniform electric fields that alternating fields of a wide range of frequencies were used. The DEP force $F_{DEP}$ acting on a spherical particle of radius $r$ suspended in a fluid of permittivity $\varepsilon_m$ is given：

$$F_{DEP} = 2\pi r^3 \varepsilon_0 \varepsilon_m \mathrm{Re}[K^*(\omega)] \nabla(E^2) \qquad (1)$$

and

$$K^*(\omega) = \frac{\varepsilon_p^* - \varepsilon_m^*}{\varepsilon_p^* + 2\varepsilon_m^*}$$

where $\varepsilon_p^*$ and $\varepsilon_m^*$ are complex permittivity of the dielectric particle and the medium, and the magnitude of the electric field, $E$, may be replaced by $E_{rms}$, which is root-mean-square of the external field, in an alternating field. $\text{Re}[K^*(\omega)]$, the real (in-phase) part of the Clausius Mossotti factor, $K^*(\omega)$, is a parameter of the effective polarizability of the particle which varies as a function of the frequency of the applied field ($\omega$), and the dielectric properties of the particle and the surrounding medium.

### III. FABRICATION OF THE MICROCHIP

The microchip of with the open-top microchannels was fabricated by microfabrication techniques, which were depicted in Fig. 2. The electrodes were made by depositing chrome and aurum sequentially on the slides by an electron beam evaporator. Chrome was the first deposited for the purpose of enhancement. The finally deposition thickness of Chrome and Aurum are 50 nm and 100 nm. The JSR photoresist (60 μm in thickness) was spun on the glass slide to pattern micro-channels of constriction for insulator-based DEP. The PDMS prepolymer mixture was cured and cut to form a well of 3 cm by 1.5 cm. The glass slide and the PDMS well were bonded after the treatment of oxygen plasma to avoid the leakage of the sample.

### IV RESULTS AND DISCUSSIONS

The contours of the square of electric field at different heights in the open-top microchip were simulated and depicted in Fig. 3. The applied electric field is $3 \times 10^4$ V/m; the space and the angle of the tip are 60 μm and 30 degree, respectively. The insulator (JSR photoresist) is 60 μm in height. On the surface and 30 μm above of the glass slide (where is the middle plane of the insulator), the local maximum of the electric fields occurred at the tips which form the constriction of the electric field. When the height increases, the electric field became uniform due to the absence of the insulator-based constriction. The distributions of the gradient of $E^2$ along the centerline between the tips at different heights (for the gap between tips 60 μm and the angle of the tip 30 degree) were plotted in Fig. 4a. The magnitude of gradient of $E^2$ is proportional to the DEP force. Comparing to the DEP force at the surface of the glass slid, the magnitude of the DEP was reduced approximately 14% and 53%, when the height was increased to 30 and 60 μm, respectively. The results in Fig 4a revealed that the DEP force decrease with the height, however, the force is still strong enough within the insulator, where the height is less than 60 μm. The shorter the distance is, the larger gradient of $E^2$ is, which implied stronger DEP force. The scanning electron microscopy (SEM) image of microfabricated trapping chip with the open-top microchannels was shown in Fig. 5. The experimental results were also performed in the present work. The experimental set up was established as shown in Fig. 6 (the tips are 100 μm apart). According to the preliminary results obtained experimentally, the 3T3 (Mouse embryonic fibroblast cell line) cells suspended in the sucrose solution (8.62 % weight percent; and its conductivity is 17.6 μS/cm) could be successfully trapped at the tip when applying the electric field of $3 \times 10^4$ V/m, and the AC frequency of 1 MHz (Fig. 7). The insulator-based DEP trapping microchip with the open-top microchannels was designed and fabricated in this work. The cells trapped by DEP force could be further treated or cultured in the open-top microchannel; however, those trapped in the microchip with enclosed microchannels could not be proceeded easily.

ACKNOWLEDGMENT

The authors would like to thank the National Science Council of the Republic of China for financial support of this research under contract No. NSC-96-2221-E-194-053.

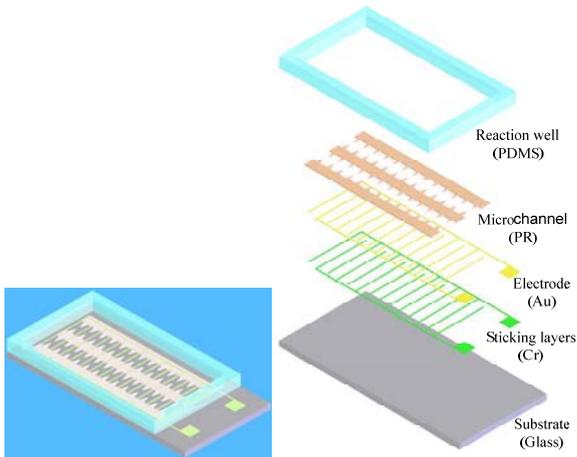

Figure 1: Schematic diagram of cell trapping chips with open-top microchannels.

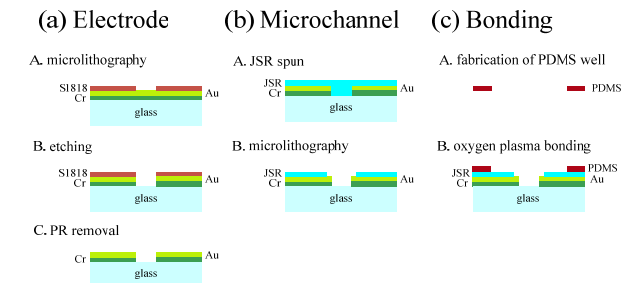

Figure 2: The fabrication processes of the cell trapping microchips: fabrication of (a) electrode on glass slide; (b) microchannel; and (c) bonding.

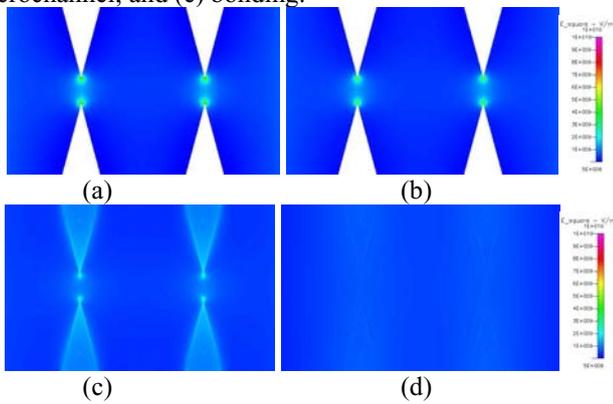

Figure 3: The numerical simulation of the electric field in the microchannel at (a) the surface of the glass slide; (b) 30 μm above the glass slide; (c) 60 μm above the glass slide (the thickness of the JSR photo resist is 60 μm); and (d) 160 μm above the glass slide. The applied electric field is 3×10$^4$ V/m and the space and the angle of the tip are 60 μm and 30 degree, respectively.

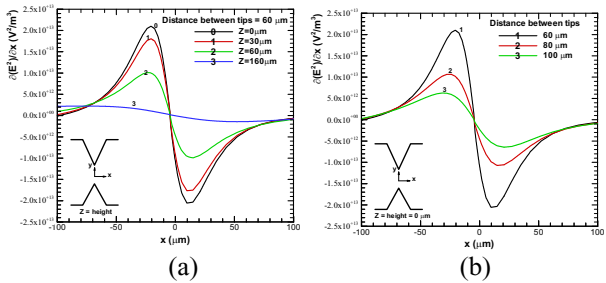

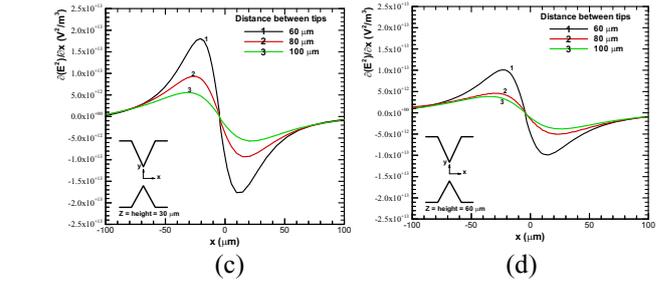

Figure 4: The numerical results of the gradient of the square of electric field along the centerline between the tips: (a) at different heights (the gap between tips 60 μm and the angle of the tip 30 degree); (b-d) for different distance between tips at heights of 0, 30 and 60 μm, respectively. The applied electric field is 3×10$^4$ V/m.

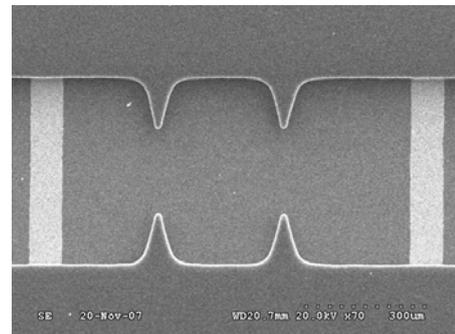

Figure 5: The scanning electron microscopy (SEM) image of the microfabricated trapping device.

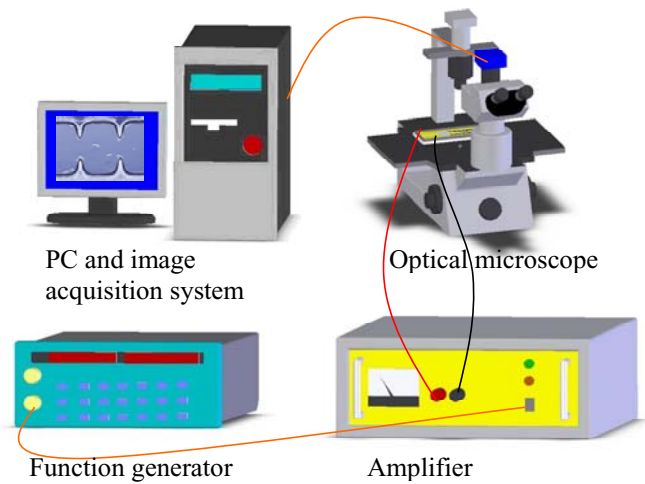

Figure 6: The schematic illustration of the experimental set up.

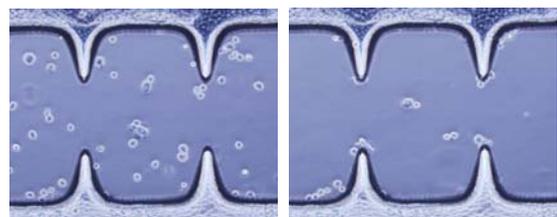

Figure 7: The experimental results of cell trapping using insulator-based DEP: (a) AC field is off; (b) AC field is on, the applied electric field is 3×10$^4$ V/m, the frequency of the AC field is 1 MHz.